\documentclass[preprint,showpacs,preprintnumbers,amsmath,amssymb]{revtex4}

\usepackage{graphicx}
\usepackage{dcolumn}
\usepackage{bm}

\begin{document}


\title{Combination of anticipated and isochronous synchronization in coupled semiconductor lasers system}

\author{Liang Wu}
\email{liangwu@suda.edu.cn}

\author{Shiqun Zhu}
\email{szhu@suda.edu.cn}

\affiliation{Department of Physics, College of Sciences, Suzhou
University, Suzhou, Jiangsu 215006, People's Republic of
China\footnote{Mailing address}
}%

\author{Yun Ni}
\affiliation{Basic Course Department, Suzhou Polytechnical
institute of agriculture, Suzhou, Jiangsu 215004 People's Republic
of China
}%


\begin{abstract}
Combination of two basic types of synchronization, anticipated and
isochronous synchronization, is investigated numerically in
coupled semiconductor lasers. Due to the combination, a
synchronization of good quality can be obtained. We study the
dependence of the lag time between two lasers and the
synchronization quality on the converse coupling retardation time
$\tau_{c21}$. When $\tau_{c21}$ is close to the difference of
external cavity round trip time $\tau$ and coupling retardation
time $\tau_{c12}$, the combination of anticipated and isochronous
synchronization may produce a better synchronization, with a lag
time proportional to $\tau_{c21}$. When $\tau_{c21}$ is largely
different from $\tau-\tau_{c12}$, the combination is noneffective
and even negative in some cases, with a lag time independent of
$\tau_{c21}$.
\end{abstract}
\maketitle

\section{Introduction}
In last few decades, much attention has been paid to chaotic
synchronization because of its potential applications in wide
variety of fields, especially in communications
\cite{Pecora90,winful90,Roy94,Sugawara94,VanWiggeren98sci,Goedgebuer98}.
Recent focus is put upon coupled semiconductor lasers operating in
Low-Frequency-Fluctuation (LFF) regime
\cite{Sivaprakasam99,Fischer00,Fujino00,Ahlers98,Mirasso96,Annovazzi-lodi96},
where the intensity output of the laser exhibits irregular dropout
events (sisyphus effect, see \cite{Heil98}). Sudden reduction in
output (a dropout) is followed by a gradual recovery and then
another dropout comes
\cite{Risch77,Heil98,Henry86,Sano94,Fischer96}. The interval
between dropouts is irregular
\cite{Hohl95,Sukow97,Marino02,Buldu04}and two lasers without
interaction to each other should exhibit severally irregular and
uncorrelated dropouts. When there is a coupling between two lasers
by injecting part output of one laser (laser1) into the other
(laser2), the two lasers may produce correlated outputs,
particularly in dropout events \cite{Wallace00}. In other words
Dropouts of laser1 is imaged into the output of laser2, thus
dropout events of two lasers appear in the same pace, and
synchronization between two lasers is achieved in this way. Such
synchronization doesn't mean always exact equality of two lasers'
outputs, but can be considered as a chaos control of laser2's LFF
behavior by laser1, in spite that complete synchronization can be
obtained in special situations
\cite{Voss00,Masoller01,Liu01apl,Tang03}.

Through a simple analysis of the rate equations for the
unidirectional coupling lasers system, two types of
synchronization has been identified
\cite{Locquet01,Locquet02ol,Locquet02pre,Shahverdiev02pla,Murakami02,Koryukin02}:
\begin{eqnarray}
Anticipated \ Synchronization (AS): \quad I_2(t-\tau)=I_1(t-\tau_{c12})\\
Isochronous \ Synchronization (IS): \quad I_2(t)=I_1(t-\tau_{c12})
\end{eqnarray}
System may choose one type of synchronization to exhibit, and the
other synchronization behavior is hidden. Which type is chosen is
determined by the competition between two types. The winner is
shown and the loser is hidden. Transition from one type to the
other may occur when operating parameters are changed
\cite{Locquet02pre,Wu03,Sivaprakasam02}. But virtually there is
not only competition but also cooperation or combination between
two types of synchronization, especially when the problem is
extended to the field of complex network with delayed feedback and
coupling \cite{Barahona02,Atay04,Albert02RMP}. In this paper,
after some brief introduction to the two basic types of
synchronization, the synchronization combination phenomenon is
discussed. To our best knowledge, this is the first paper to
discuss the combination of the two basic types of chaotic
synchronization in coupled semiconductor lasers system.

\section{Anticipated and Isochronous Synchronization}
The two basic types of synchronization are reviewed in this
section.

\begin{figure}
\includegraphics{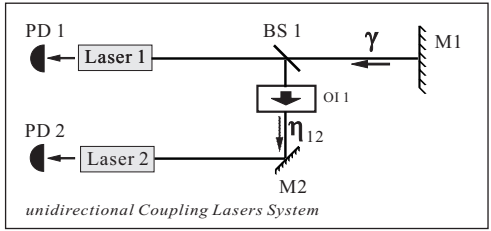}
\caption{\label{fig:epsart}Schematic representation of
unidirectional coupling lasers system, where two basic types of
synchronization are shown typically.}
\end{figure}

The unidirectional coupling lasers system is shown in Fig.1. Part
output of laser1 is injected back as a feedback by a mirror M1.
Another part is injected into laser2 via a beam splitter BS1, an
optical isolator OI1 (ensure there is no light from laser2 to
enter laser1 and alter the dynamics of laser1) and another mirror
M2. No feedback is used in laser2. Photo diodes (PD1, 2) is used
to detect outputs of two lasers respectively. Feedback rate is
labelled by $\gamma$, $\eta_{12}$ is coupling strength from laser1
to laser2.

Numerical simulation is performed by LK equations for the complex
electric fields $E$ and normalized carrier densities $N$
\cite{Lang80}.
\begin{eqnarray}
\frac{dE_1}{dt} & = &
k(1+i\alpha)[G_1-1]E_1(t) +\gamma_1 E_1(t-\tau)e^{(-i\omega_1\tau)}+\beta\xi_1(t)\\
\frac{dN_1}{dt} & = & \frac{j-N_1-G_1|E_1|^2}{\tau_{n}}\\
\frac{dE_2}{dt} & = & k(1+i\alpha)[G_2-1]E_2(t)+\eta_{12}E_1(t-\tau_{c12})e^{[-i(\omega_1 \tau_{c12}+\triangle\omega t)]}+\beta\xi_2(t)\\
\frac{dN_2}{dt} & = & \frac{j-N_2-G_2|E_2|^2}{\tau_{n}}
\end{eqnarray}
Where subscripts 1 and 2 denote \textit{Laser1} and
\textit{Laser2} respectively. The second term in Eq. (1)
corresponds to the feedback in \textit{Laser1}, and the second
term in Eq. (3) corresponds to the coupling from laser1 to laser2.
$k$ is the cavity loss, $\alpha$ the linewidth enhancement factor,
$G=N/(1+\epsilon|E|^2)$ is the optical gain, $\epsilon$ is the
gain saturation coefficient, $\omega$ is the optical frequency
without feedback, $\triangle\omega=\omega_2-\omega_1$ is the
frequency detuning, $\xi$ is independent complex Gaussian white
noise, and $\beta$ measures the noise strength, $j$ is the
normalized injection current, and $\tau_n$ is the carrier
lifetime. For simplicity, we may choose $\beta_1=\beta_2= 0$. Two
types of synchronization are shown in following two cases
respectively.

\begin{figure}
\includegraphics{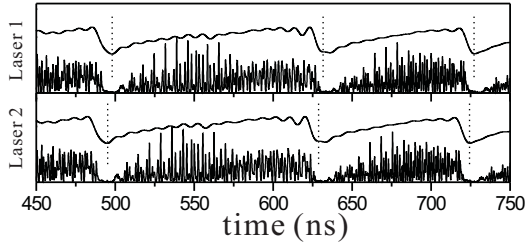}
\caption{\label{fig:epsart}Time traces of two unidirectional
coupling lasers operating in the LFF regime, typically showing
anticipated synchronization with parameters:
$\eta_{12}=\gamma=8ns^{-1}$, $\tau=7ns$, $\tau_{c12}=4ns$,
$j=1.003$, $\alpha=5$, $k=500ns^{-1}$, $\tau_n=1ns$. Vertically
shifted low-pass-filtered ones are plotted in solid lines to
emulate the experimental detection and exhibit the dropout events
clearly \cite{Fischer96}.}
\end{figure}

In the first case, a typical AS (anticipated synchronization) is
shown with parameters: $\eta_{12}=\gamma=8ns^{-1}$, $\tau=7ns$
($\tau$ is external cavity round trip time in laser1),
$\tau_{c12}=4ns$ ($\tau_{c12}$ is the time for light to travel
from laser1 to laser2). The intensity outputs of two lasers are
plotted in Fig. 2. The two lasers are both operating in LFF
regime, where sudden output reduction (a dropout) appears,
followed by a gradual recovery, and then another dropout comes.
Intervals between dropouts is irregular. To remove the
fluctuations in high frequencies so as to make the dropout events
clearer, the low-pass-flitter diagrams are also plotted and
shifted upward. The dropout events is labelled by short dotted
lines.

As clearly shown in Fig. 2, the two lasers undergo dropouts in the
same pace due to the coupling and synchronization is achieved. In
addition, 3ns before every dropout in laser1, there is always a
corresponding dropout in laser2, as if laser2 can predict the
future state of laser1 in spite that laser2's chaotic dynamics is
driven by laser1. This is a typical AS, satisfying Eq. (1).

\begin{figure}
\includegraphics{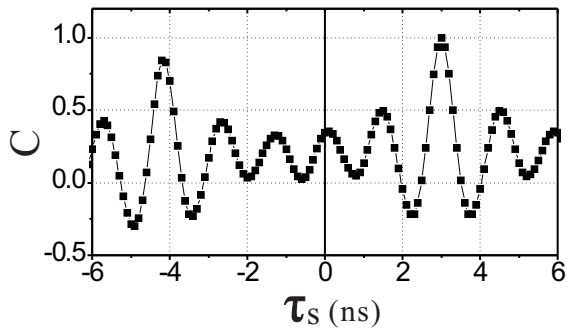}
\caption{\label{fig:epsart}plot of correlation function C as a
function of shift time $\tau_s$, corresponding to the typical
anticipated synchronization in Fig. 2.}
\end{figure}

In Fig. 3 correlation function C between two lasers' outputs as a
function of shift time ($\tau_s$) is also plotted. C is defined as
follows:
\begin{eqnarray}
C(\tau_s)=\frac{\langle[I_1(t+\tau_s)-\langle
I_1\rangle][I_2(t)-\langle I_2\rangle]\rangle}{\{\langle
[I_1(t)-\langle I_1\rangle]^2\rangle\langle [I_2(t)-\langle
I_2\rangle]^2\rangle\}^{1/2}}
\end{eqnarray}
At $\tau_s=3ns$ there is a peak. This peak indicates that a strong
correlation can be obtained when laser2 is shifted backward by
3ns. Obviously this peak corresponds to the AS in Fig. 2.

Besides the right peak, there is another peak at $\tau_s\approx
-4ns$. (It is noted that $\tau_{c12}=4ns$ in our simulation. This
peak indicates that shifting laser2 forward by $\tau_{c12}$ also
can produce a strong correlation. This left peak satisfies Eq. (1)
and corresponds to IS. Although only AS is shown in Fig. 2,
two-peaks configuration in correlation plot implies the existence
of IS. Only because the right peak is higher (indicating that the
anticipated synchronization quality is better), AS is the winner
in the competition with IS, and IS behavior is hidden.

\begin{figure}
\includegraphics{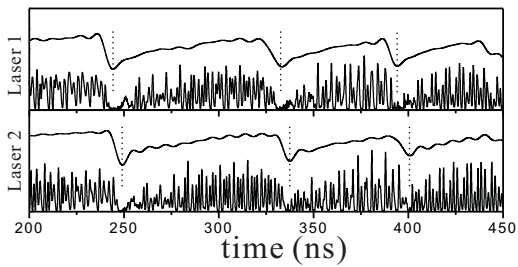}
\caption{\label{fig:epsart}Time traces of two unidirectional
coupling lasers typically showing isochronous synchronization with
parameters: $\eta_{12}=5ns^{-1}$, other parameters are the same as
in Fig. 2.}
\end{figure}

In the second case, a typical IS (isochronous synchronization) is
shown with parameters: $\eta_{12}=8ns^{-1}$, $\gamma=5ns^{-1}$,
$\tau=7ns$, $\tau_{c12}=4ns$, namely only $\gamma$ is decreased a
little. Outputs of two lasers and vertically-shifted
low-pass-filtered diagrams are plotted in Fig. 4, ranging from
200ns to 450ns. Every dropout in laser1 leads the corresponding
dropout by about 4ns (We note $\tau_{c12}=4ns$). So there is a
typical IS between the two lasers, satisfying Eq. (2). This
synchronization is induced by the fact that the unidirectional
coupling is sufficient for the laser1 to drive the dynamics of the
laser2 and lead to a locking-state phenomenon.

\begin{figure}
\includegraphics{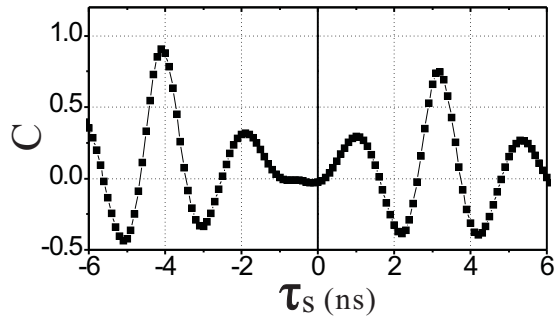}
\caption{\label{fig:epsart}plot of correlation function C as a
function of shift time $\tau_s$, corresponding to the typical
isochronous synchronization in Fig. 4. }
\end{figure}

Fig. 5 is the correlation plot. It is very easy to identify the
two peaks. The left peak corresponds to IS and the right one
corresponds to AS. Only because the left peak is higher than the
right one (indicating that the isochronous synchronization quality
is better), AS is the loser in the competition with IS and has
been hidden. Hence it is clearly seen in Fig. 4 that only IS is
shown.

According to above discussion, which type of synchronization is
shown, AS or IS, is determined by their competition. The winner's
behavior is shown, while the loser still exists in the system but
has been hidden.

\section{Synchronization Combination}

\begin{figure}
\includegraphics{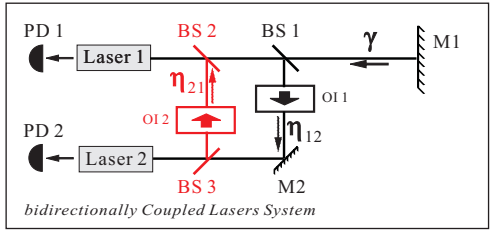}
\caption{\label{fig:epsart}Schematic representation of
bidirectional coupling lasers system, where the combination of
anticipated and isochronous synchronization is investigated. The
coupling from laser1 to laser2 travels via BS1, OI1 and M2, while
the converse coupling from laser2 to laser1 travels via BS3, OI2
and BS2.}
\end{figure}

AS and IS not only compete but also cooperate with each other,
i.e. their combination. In this section such combination is
discussed. the model under our following discussion is almost the
same as the second case in the previous section except that a
converse coupling with coupling strength $\eta_{21}$ is used as
shown in Fig. 6 with parameters: $\eta_{21}=\eta_{12}=8ns^{-1}$,
$\gamma=5ns^{-1}$. Two optical isolators are used to make sure
that both couplings are unidirectional, then
$\tau_{c21}\neq\tau_{c12}$, which is helpful in following
analysis. $\tau_{c21}$ is the time taken for the light to travel
from laser2 back to laser1.

Then the right correlation peak appears because of a combination
of two factors:

(1)Before adding the converse coupling, there has been an AS:
$I_2(t-\tau)=I_1(t-\tau_{c12})$, where laser2 lags laser1 by the
time $\tau_s=\tau-\tau_{c12}$.

(2)After adding the converse coupling, there is also an IS due to
the converse coupling from laser2 to laser1:
$I_1(t)=I_2(t-\tau_{c21})$, where laser2 lags laser1 by the time
$\tau_s=\tau_{c21}$.

We may choose: $\tau=7ns$, $\tau_{c12}=4ns$, and $\tau_{c21}=3ns$,
then $\tau_{c21}=\tau-\tau_{c12}$, and there will be a combination
of AS with IS at $\tau_s=3ns$.

The dynamical behavior of bidirectional coupling laser1 (with a
external feedback)and laser2 (solitary) is described by Eq. (8, 9)
and Eq. (5, 6) respectively.
\begin{eqnarray}
\frac{dE_1}{dt} & = & k(1+i\alpha)[G_1-1]E_1(t) +\gamma_1 E_1(t-\tau)e^{(-i\omega_1\tau)}+\eta_{21}E_2(t-\tau_{c21})e^{[-i(\omega_2 \tau_{c21}-\triangle\omega t)]}+\beta\xi_1(t)\\
\frac{dN_1}{dt} & = & \frac{j-N_1-G_1|E_1|^2}{\tau_{n}}
\end{eqnarray}
where the third term in Eq. (8) is added for the converse
coupling.

\begin{figure}
\includegraphics{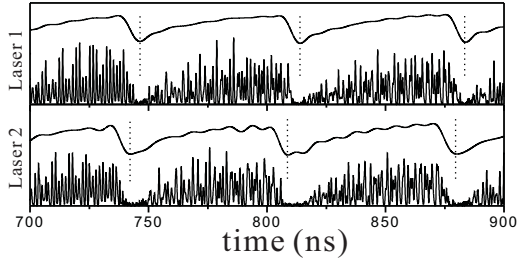}
\caption{\label{fig:epsart} Time traces of two lasers, showing the
effect of the combination of anticipated and isochronous
synchronization, laser2 is seen to synchronize and lead laser1.
parameters are: $\eta_{21}=8ns^{-1}$, $\tau_{c21}=3ns$, and other
parameters are the same as in Fig. 4.}
\end{figure}

\begin{figure}
\includegraphics{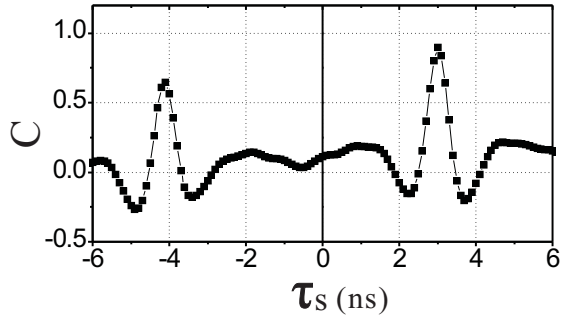}
\caption{\label{fig:epsart}plot of correlation function C as a
function of shift time $\tau_s$, corresponding to the combination
of anticipated and isochronous synchronization in Fig. 7.}
\end{figure}

Numerical simulation is shown in Figs. 7 and 8. As expected, in
Fig. (7) 3ns before every dropouts of laser1 there has been a
dropout in laser2 inevitably. Synchronization between two lasers
is achieved and Laser2 leads laser1 by 3ns.

That Laser2 can predict the future dynamical behavior of laser1
results from the combination of two basic types of synchronization
behavior, AS and IS. The effect of combination is also exhibited
by the correlation plot Fig. 8. The right correlation peak at
$\tau_s=3ns$, reflecting the effect of combination of AS and IS,
is higher. While the left peak at $\tau_s\approx -4ns$ reflecting
IS produced by only the coupling from laser1 to laser2 is lower
and its corresponding dynamical behavior is hidden. So only the
\emph{combination synchronization} is shown.

It seems that the matching condition $\tau_{c21}=\tau-\tau_{c12}$
makes the combination of AS and IS create a strong correlation and
consequent \emph{combination synchronization}. For further
discussion about the effect of combination of AS and IS, a more
general situation where $\tau_{c21}\neq \tau-\tau_{c12}$ is
concerned by moving the branch of the converse coupling so that
$\tau_{c21}$ is modified.

In the following discussion, combination synchronization quality
$Q_c$ and its characteristic time scale $T_c$ are investigated:
\begin{eqnarray}
Q_c  =  \max(C(\tau_s)), \quad \tau_s>0\\
C(T_c))  =  Q_c
\end{eqnarray}
where $T_c$ is the lag time between two lasers.

(1) In AS, laser1 lags laser2 by the time dependent on the
difference of the external cavity round trip time of laser1 and
the coupling retardation time , i.e. $\tau-\tau_{c12}$,
independent on the converse coupling retardation time
$\tau_{c21}$. When $\tau_{c21}$ is increased, the lag time should
almost remain unchanged.

(2) In IS, laser1 lags laser2 by the time dependent on the
converse coupling retardation time $\tau_{c21}$. When $\tau_{c21}$
is increased, the lag time should be increased proportionally.

In combination of AS and IS, whether the lag between two lasers
($T_c$) is changed proportionally or unchanged when $\tau_{c21}$
is increased?

\begin{figure}
\includegraphics{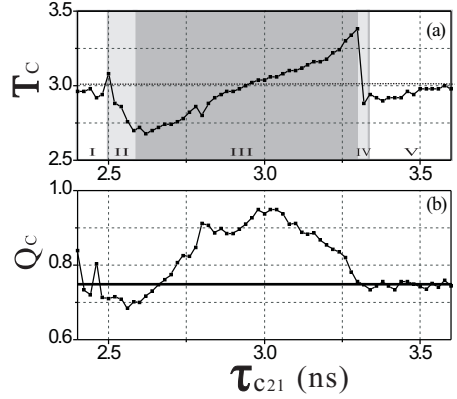}
\caption{\label{fig:epsart}shows the dependence of $T_c$ and $Q_c$
on the converse coupling retardation time $\tau_{c21}$ ranged from
2.4ns to 3.6ns. Five regions are identified. All the parameters
are the same as in Fig. 7 except for the variable $\tau_{c21}$.}
\end{figure}

Fig. 9 shows the dependence of $T_c$ (a) and combination
synchronization quality $Q_c$ (b) on the converse coupling
retardation time $\tau_{c21}$. The interesting results show the
range of $\tau_{c21}$ from 2.4ns to 3.6ns can be divided into five
regions, labelled by I-V. In region I and V, $T_c$ is unchanged
and always stay near 3.0ns, that indicates when $\tau-\tau_{c12}$
and $\tau_{c21}$ is largely different as in region I and V, AS
behavior is exhibited and more prominent in the combination of AS
and IS. Furthermore, it is seen in Fig. 9(b) that corresponding
$Q_c$ is near the quality of only AS (The quality of only AS is
shown by the height of the right peak in Fig. 5, where the
converse coupling is still not added, and is also lined out in
Fig. 9(b) by a straight line.) It seems that when two
characteristic time scales $\tau-\tau_{c12}$ and $\tau_{c21}$ are
largely different, The combination effect is not so great to
enhance the synchronization quality. In region III where
$\tau_{c21}$ is close to $\tau-\tau_{c12}$, $T_c$ always
approximately equals to $\tau_{c21}$ ($T_c\approx \tau_{c21}$),
presenting a striking contrast with Region I and V. The fact that
$T_c$ increases with $\tau_{c21}$ in region III implies IS
behavior is prominent and the decisive force in the combination of
AS and IS. In addition, the quality of the combination
synchronization goes over the straight line mostly, indicating the
combination of IS and AS has produced a better synchronization
than only AS in most of region III. The maximum efficiency of the
combination is obtained at about $\tau_{c21}=3ns$
($\tau-\tau_{c12}=3ns$). Regions II and IV are transition regions.
$T_c$ falls rapidly from $\tau-\tau_{c12}$ to $\tau_{c21}$ in
region II, and from $\tau_{c21}$ back to $\tau-\tau_{c12}$ in
region IV. In this transition regions the situation that neither
AS nor IS is prominent in the combination may occur, where the
quality of the resulting combination synchronization is even worse
than only AS, especially in region II.

\section{Discussion}
We note that In Fig. 7 where the combination synchronization is
demonstrated, the both couplings are of the same strength, i.e.
$\eta_{12}=\eta_{21}$. If the feedback in laser1 is removed, the
system will turn to be a typical Face-to-Face (F2F) model
\cite{Heil01,Mulet02}. In F2F model, synchronization can be
obtained with a time lag between two lasers. And the leader role
switches from one laser to the other randomly and continuously
because of the symmetry. Without the explanation from the
viewpoint of synchronization combination, it is very hard to
understand why laser2 will always leads laser1 after a delayed
feedback is added in laser1.

A simple analysis of the rate equations for the unidirectional
lasers system has identified the required matching condition
$\eta_{12}=\gamma$ (no feedback in laser2) necessary for the
existence of AS. We note this condition is necessary for the
complete synchronization. In fact, when the stringent matching
condition is not satisfied AS still exists, but it is not so good
and always hidden behind. As shown in Fig. 7, such not so good AS
has even become an important factor in the synchronization
combination.

Sivaprakasam \textit{et al} experimentally demonstrated a very
interesting synchronization phenomenon between two bidirectional
coupling lasers \cite{Sivaprakasam01,Rees03}. Their setup
configuration is the same as Fig. 6 of this paper except for
$\tau_{c12}=\tau_{c21}$ in their setup. They found experimentally
that "slave" can predict the future state of the "master" and the
corresponding "anticipating time" always equals to the coupling
retardation time. We thind the experimentally obtained
synchronization may arise from the combination of synchronization.
And in region III of Fig. 9 of this paper, $T_c=\tau_{c21}$ is in
good agreement with their experimental result.

\bibliography{Combination}
\end{document}